\documentclass[twocolumn]{aastex62}
\usepackage{natbib}
\usepackage{amsmath}
\usepackage{floatrow}
\published{AJ}

\shorttitle{Sample article}
\shortauthors{Morello et al.}

\begin{document}

\title{The ExoTETHyS package: Tools for Exoplanetary Transits around Host Stars}

\correspondingauthor{Giuseppe Morello}
\email{giuseppe.morello@cea.fr}

\author[0000-0002-4262-5661]{G. Morello}
\affil{AIM, CEA, CNRS, Universit\'e Paris-Saclay, Universit\'e Paris Diderot, Sorbonne Paris Cit\'e, F-91191 Gif-sur-Yvette, France}
\affil{INAF--Osservatorio Astronomico di Palermo, Piazza del Parlamento 1, 90134 Palermo, Italy}
\affil{Dept. of Physics \& Astronomy, University College London, Gower Street, WC1E 6BT London, UK}

\author{A. Claret}
\affil{Instituto de Astrof\'isica de Andaluc\'ia, CSIC, Apartado 3004, 18080 Granada, Spain}
\affil{Dept. F\'{\i}sica Te\'{o}rica y del Cosmos, Universidad de Granada, Campus de Fuentenueva s/n,  10871, Granada, Spain}

\author{M. Martin-Lagarde}
\affil{AIM, CEA, CNRS, Universit\'e Paris-Saclay, Universit\'e Paris Diderot, Sorbonne Paris Cit\'e, F-91191 Gif-sur-Yvette, France}

\author{C. Cossou}
\affil{Institut d'Astrophysique Spatiale, CNRS/Universit\'e Paris-Sud, Universit\'e Paris-Saclay, b\^atiment 121, Universit\'e Paris-Sud, 91405 Orsay Cedex, France}

\author{A. Tsiaras}
\affil{Dept. of Physics \& Astronomy, University College London, Gower Street, WC1E 6BT London, UK}

\author{P.-O. Lagage}
\affil{AIM, CEA, CNRS, Universit\'e Paris-Saclay, Universit\'e Paris Diderot, Sorbonne Paris Cit\'e, F-91191 Gif-sur-Yvette, France}



\begin{abstract}
\noindent
We present here the first release of the open-source python package \texttt{ExoTETHyS} (stable: \url{https://zenodo.org/badge/latestdoi/169268509}, development version: \url{https://github.com/ucl-exoplanets/ExoTETHyS/}), which  aims to provide a stand-alone set of tools for modeling spectrophotometric observations of the transiting exoplanets. In particular, we describe: (1) a new calculator of stellar limb-darkening coefficients that outperforms the existing software by one order of magnitude in terms of light curve model accuracy, i.e., down to $<$10 parts per million, and (2) an exact transit light curve generator based on the entire stellar intensity profile rather than limb-darkening coefficients. New tools will be added in later releases to model various effects in exoplanetary transits and eclipsing binaries. \texttt{ExoTETHyS} is a reference package for high-precision exoplanet atmospheric spectroscopy with the upcoming \emph{James Webb Space Telescope} and \emph{Atmospheric Remote-sensing Infrared Exoplanet Large-survey} missions.
\vspace{0.2 cm}

\noindent
\emph{Unified Astronomy Thesaurus concepts:} Observational astronomy (1145); Spectroscopy (1558); High time
resolution astrophysics (740); Exoplanet atmospheres (487); Stellar atmospheres (1584); Transit photometry
(1709); Eclipsing binary stars (444); Exoplanet systems (484); Stellar astronomy (1583); Limb darkening (922)
\vspace{0.2 cm}

\noindent
Software reviewed by the \href{https://doi.org/10.21105.joss.01834}{Journal of Open Source Software}

\end{abstract}


\section{Introduction} \label{sec:intro}
\begin{deluxetable*}{ccccccc}[t!]
\tablecaption{Stellar model atmosphere grids available with the first release of ExoTETHyS  \label{tab:stellar_grids}}
\tablecolumns{6}
\tablenum{1}
\tablewidth{0pt}
\tablehead{
\colhead{Name} & \colhead{Geometry\tablenotemark{a}} &
\colhead{Range $T_{\mbox{eff}}$(K)} &
\colhead{Range $\log{g}$} & \colhead{Range $[M/H]$} & \colhead{Range $\lambda$ ($\mu$m)} & \colhead{Reference}
}
\startdata
\texttt{ATLAS} & P-P & 3500-50000 & 0.0-5.0 & --5.0-1.0 & 0.009-160.0 & \citet{claret00} \\
\texttt{PHOENIX}\_2012\_13 & S1 & 3000-10000 & 0.0-6.0 & 0.0 & 0.25-10.0 & \citet{claret12,claret13} \\
\texttt{PHOENIX}\_2018 & S1 & 2300-12000 & 0.0-6.0 & 0.0 & 0.05-2.6 & \citet{claret18} \\
\enddata
\tablenotetext{a}{Geometry types: P-P=plane-parallel; S1=spherical 1D}
\end{deluxetable*}
More than 3000 transiting exoplanets have been discovered in the last 20 years. The number of transiting exoplanets accounts for about three-quarters of the current exoplanet census\footnote[7]{source: \url{https://exoplanetarchive.ipac.caltech.edu}}, although this large fraction is due to targeted research programs rather than being a random sample from the exoplanet population. The success of the transit method is due to several contributing factors, including its ability to characterize them in great detail.
A transit is revealed by a decrement in flux while the planet occults part of the stellar disk. The main observables are the transit depth and durations, leading to measurements of the exoplanet size, orbital semimajor axis and inclination, and stellar mean density \citep{seager03}. Transit spectroscopy is now routinely used to investigate the chemistry and physics of exoplanet atmospheres, through differences in transit depth of $\sim$10-100 parts per million (ppm) relative to the stellar flux at multiple wavelengths (e.g., \citealp{iyer16, sing16, tsiaras18}).

Accurate modeling of the host star effects is mandatory to achieve the spectrophotometric precision required for characterizing the atmosphere of transiting exoplanets. The most prominent effect is stellar limb-darkening \citep{mandel02}, followed by magnetic activity \citep{ballerini12, zellem17}, granulation \citep{chiavassa17}, and, in some cases, rotational oblateness and gravity darkening \citep{howarth17}, and tidal deformations \citep{akinsanmi19, hellard19}. Among the nonstellar effects, the exoplanet nightside emission can also play a significant role \citep{kipping10, morello19}.

The \texttt{ExoTETHyS} package is conceived as a toolbox for those who analyze the exoplanetary transits. The first release focuses on the tools for modeling the stellar limb-darkening effect, the importance of which is ubiquitous in transit observations, as well as in optical interferometry, microlensing, and eclipsing binary observations.
Future versions of \texttt{ExoTETHyS} will include useful tools for modeling other effects, as well as for estimating their impact on specific observations, based on the astrophysical system parameters, the instrument passband, and the noise level. Accurate modeling of all of the aforementioned effects proved to be crucial in the analysis of several \emph{CoRoT} and \emph{Kepler} objects (e.g., \citealp{mazeh10, barnes11, mazeh12,  masuda15, howarth17, reinhold17, shporer17, nielsen19}), because of the high-precision photometry down to the $\lesssim$10~ppm level \citep{christiansen12}. A similar photometric precision is expected for some of the ongoing \emph{Transiting Exoplanet Survey Satellite} (\emph{TESS}) observations \citep{ricker14}, future observations with the \emph{CHaracterising ExOPlanet Satellite} (\emph{CHEOPS}; \citealp{isaak19}) and \emph{PLAnetary Transits and Oscillations} (\emph{PLATO}; \citealp{rauer14}), and in spectroscopic observations with the upcoming \emph{James Webb Space Telescope} (\emph{JWST}; \citealp{beichman14}) and \emph{Atmospheric Remote-sensing Infrared Exoplanet Large-survey} (\emph{ARIEL}; \citealp{pascale18}) space missions.

Stellar limb-darkening is the wavelength-dependent radial decrease in specific intensity. Consequently, the transit light curve deviates from the flat-bottomed  shape that would be observed in the case of a uniform stellar disk; the difference signal can be as large as $\sim$10$^4$ ppm for the transit of a hot Jupiter observed at UV or visible wavelengths.
Typically, the radial intensity distribution computed from specific stellar atmosphere models is parameterized by a set of limb-darkening coefficients, which are fixed in the analyses of transit light curves. Many researchers have produced multiple grids of stellar atmosphere models with different codes, then used to compile precalculated tables of limb-darkening coefficients (e.g., \citealp{claret00, claret03, claret04, claret08, claret17, claret18, sing10, howarth11b, claret11, claret12, claret13, claret14, neilson13, neilson13b, magic15, reeve16}). The lack of empirical validation for stellar limb-darkening prevents the final choice of the most reliable model(s). The presence of unocculted stellar spots during an exoplanetary transit may alter the effective limb-darkening coefficients, which will be slightly different from those calculated for the case of unspotted stellar surface \citep{csizmadia13}.

In some cases, significantly different parametric intensity profiles have been obtained from the same model atmosphere, depending on the sampling of the model intensity profile, the functional form (so-called limb-darkening law), and/or the fitting algorithm adopted \citep{claret00, heyrovsky07, howarth11, espinoza15}. The system parameters obtained from the light curve fits with the alternative sets of limb-darkening coefficients can vary by more than the respective 1$\sigma$ error bars, typically, if the relative photometric precision of the observations is of the order of (or better than) 100~ppm per minute interval.

In this paper, we probe an optimized fitting algorithm for the limb-darkening coefficients that minimizes the difference between (numerically integrated) reference light curves and the corresponding approximated transit models with limb-darkening coefficients. Therefore we eliminate the degeneracy from the choice between several fitting algorithms that were leading to significantly different parametric profiles for the same stellar atmosphere model (e.g., \citealp{espinoza15}). The high-fidelity match between the stellar intensity profiles and the transit light curve models facilitates comparative studies of the model atmospheres, especially with the increasing number of observations with a spectrophotometric precision down to $\sim$10~ppm (e.g., \emph{CoRoT}, \emph{Kepler}, \emph{TESS}, and  \emph{Hubble Space Telescope} (\emph{HST})/WFC3 data).

\subsection{Structure of the paper}
Section~\ref{sec:exotethys} provides a technical description of the \texttt{ExoTETHyS} package and the algorithms adopted. Section~\ref{sec:performances} discusses the precision of the limb-darkening calculator for the analysis of exoplanetary transits. In particular, Section~\ref{ssec:fitting_methods} compares various algorithms that are adopted in the other publicly available codes and their variants, Section~\ref{ssec:fitting_ld_laws} compares the performances of the alternative limb-darkening laws, and Section~\ref{ssec:GOF_ppm} provides a formula to estimate the potential error in the transit model based on the goodness of fit for the limb-darkening coefficients that should be compared with the noise level in the observations. Section~\ref{sec:usage} discusses the main functionality of the \texttt{ExoTETHyS} package, its current and future usage. Finally, Section~\ref{sec:conclusions} summarizes the key points discussed in this paper.

\section{Description of the \texttt{ExoTETHyS} package} \label{sec:exotethys}

The first release of \texttt{ExoTETHyS} includes the following subpackages:
\begin{enumerate}
\item Stellar Atmosphere Intensity Limb (SAIL), which can calculate the limb-darkening coefficients for specific stellar targets or over predetermined parameter grids;
\item Transit Ring-Integrated Profile (TRIP), which can compute an exact transit light curve by direct integration of the occulted stellar flux, without using an analytical function (limb-darkening law) to approximate the stellar intensity profile.
\end{enumerate}
The TRIP subpackage was conceived to model exoplanetary transits. Following requests by users, we are adding a function to model eclipsing binaries.

\subsection{The SAIL subpackage}
The SAIL subpackage is a generic stellar limb-darkening calculator that is not specific to a predetermined list of instruments or standard passbands. It is conceptually similar to the calculator provided by \cite{espinoza15}, but with different features.
A technical difference is the use of a novel fitting algorithm for obtaining the limb-darkening coefficients, specifically optimized for modeling the exoplanetary transits, instead of multiple algorithm options with unclear performances (see Sections~\ref{ssec:fit_ints} and \ref{ssec:fitting_methods}). 

\subsubsection{Input and output}
\label{ssec:sail_IO}
The SAIL subpackage requires a configuration file to specify the desired calculation. The user can choose either ``individual'' or ``grid'' calculation type. The first option enables calculation of the limb-darkening coefficients for a star or for a list of stars with the parameters specified by the user, while the latter will provide the limb-darkening coefficients for a grid of precalculated stellar model atmospheres. In both cases, the user must select one of the available stellar model grids, which were computed with different codes and settings (see Table~\ref{tab:stellar_grids} and references therein).
For each grid, the stellar models are identified by a set of three parameters, i.e., the effective temperature ($T_{\mbox{\footnotesize{eff}}}$), the surface gravity ($\log{g}$), and the metallicity ($[M/H]$). As the limb-darkening coefficients are mostly dependent on the effective temperature, the user must provide the effective temperatures of all the individual stars. The other parameters have default values of $\log{g}=$4.5 and $[M/H]=$0.0, corresponding to a main-sequence star with solar abundances, if they are not given by the user. For the grid calculation type, the default option is to calculate the limb-darkening coefficients for all the stellar models in the selected database. Alternatively, the user can select a subgrid by specifying the minimum and/or maximum values for each stellar parameter.

Another key input is the passband, i.e., the total spectral response of the observing instrument. For most instruments, the spectral response is available as a table of photon-to-electron conversion factors at given wavelengths. The limb-darkening coefficients do not depend on the absolute values of the spectral response, so that a scaled/normalized version of the spectral response will give identical results. The spectral responses of the most common instruments for transiting exoplanets are built into the package. The code can accept any user-defined passband with the same file format. It is also possible to calculate the limb-darkening coefficients for multiple wavelength bins within a given passband by specifying the two wavelengths delimiting each bin. This option is particularly useful for exoplanet spectroscopic observations, such as those currently performed with \textit{HST}/WFC3.

The last mandatory input in the configuration file is the list of limb-darkening laws to adopt (at least one). The code includes several built-in limb-darkening laws, including all of the most commonly used (see Section~\ref{ssec:ld_laws}), but it can also accept user-defined laws.

The ``basic'' outputs are python dictionaries containing the best-fit limb-darkening coefficients obtained for the required passbands, wavelength bins, and limb-darkening laws. The output dictionaries also provide the corresponding weighted rms of the fitting residuals to allow for a quick quality check (see Section~\ref{ssec:GOF_ppm}). For the case of individual calculation type, the results obtained for each target are stored in separate pickle files. Optionally, the user can request a ``complete'' output, whic includes intermediate products such as the numeric intensity profiles at various stages of the calculation (see Sections~\ref{ssec:integ_ints}-\ref{ssec:ldc_interp}). The additional information of the complete output is offered, mainly, as a way to identify bugs in the code and/or issues with certain stellar model atmospheres and wavelengths. Usually, the exoplanetary scientists will be interested to the basic output only.

\subsubsection{From the stellar model atmospheres to the passband-integrated intensities}
\label{ssec:integ_ints}
The stellar model atmosphere grids consist of one file for each triple of stellar parameters ($T_{\mbox{\footnotesize{eff}}}$, $\log{g}$, $[M/H]$), providing the specific intensities ($I_{\lambda}(\mu)$) in units of erg cm$^{-2}$ s$^{-1}$ \AA$^{-1}$ sr$^{-1}$ at several positions on the sky-projected stellar disk over a given spectral range. For historical reasons, the independent variable is $\mu=\cos{\theta}$, where $\theta$ is the angle between the line of sight and the corresponding surface normal. The radial coordinate in the sky-projected disk is $r=\sqrt{1-\mu^2}$, where $r=1$ ($\mu=0$) corresponds to the spherical surface radius.
Table~\ref{tab:stellar_grids} reports the information about the databases available with the first release of ExoTETHyS. We refer to the relevant papers and references therein for comparisons between the models. The passband-integrated intensities are calculated as
\begin{equation}
\label{eqn:integrated_intensities}
I_{\mbox{\footnotesize pass}} (\mu) \propto \int_{\lambda_1}^{\lambda_2} I_{\lambda}(\mu) R_{\mbox{\footnotesize pass}}(\lambda) \lambda d \lambda ,
\end{equation}
where $R_{\mbox{\footnotesize pass}}(\lambda)$ is the spectral response of the instrument in electrons photon$^{-1}$, and $\lambda_1$ and $\lambda_2$ are the passband or wavelength bin limits. The passband-integrated intensities are obtained in units proportional to electrons cm$^{-2}$ s$^{-1}$ sr$^{-1}$. As the limb-darkening coefficients are not affected by the (omitted) proportionality factor in Equation~\ref{eqn:integrated_intensities}, the final intensities are normalized such that $I_{\mbox{\footnotesize pass}} (\mu=0) = 1$.

The intensity profiles, $I_{\lambda}(\mu)$, have distinctive behaviors depending on the plane-parallel or spherical geometry adopted by the selected grid of model atmospheres. In particular, the spherical intensity profiles show a steep drop-off close to the stellar limb, which is not observed in the plane-parallel models. The explanation for the different behaviors is exhaustive in the literature \citep{wittkowski04, espinoza15, morello17}. 
The almost null intensities at small $\mu$ are integrated over lines of sight that intersect only the outermost atmospheric shells, which have the smallest emissivity. Here $\mu=$0 ($r=$1) corresponds to the outermost shell of the model atmosphere, which is typically outside the stellar radius that would be observed in transit. Our algorithm calculates the photometric radius at the inflection point of the spherical intensity profile, i.e., where the gradient $|dI(r)/dr|$ is the maximum \citep{wittkowski04, espinoza15}. The radial coordinates are then rescaled such that $r=1$ ($\mu=$0) at the photometric radius, and those intensities with rescaled $r>$1 are rejected. No rescaling is performed for the plane-parallel models.

\subsubsection{Limb-darkening laws} \label{ssec:ld_laws}
A long list of analytical forms, so-called limb-darkening laws, has been proposed in the literature to approximate the stellar intensity profiles. The following options are built in the package:
\begin{enumerate}
\item the linear law \citep{schwarzschild06},
\begin{equation}
\label{eqn:ld_law_linear}
I_{\lambda}(\mu) = 1 - a(1-\mu) ;
\end{equation}
\item the quadratic law \citep{kopal50},
\begin{equation}
\label{eqn:ld_law_quadratic}
I_{\lambda}(\mu) = 1 - u_1(1-\mu) - u_2(1-\mu)^2 ;
\end{equation}
\item the square-root law \citep{diaz-cordoves92},
\begin{equation}
\label{eqn:ld_law_sqrt}
I_{\lambda}(\mu) = 1 - v_1(1-\sqrt{\mu}) - v_2(1-\mu) ;
\end{equation}
\item the power-2 law \citep{hestroffer97},
\begin{equation}
\label{eqn:ld_law_power2}
I_{\lambda}(\mu) = 1 - c(1-\mu^{\alpha}) ;
\end{equation}
\item the four-coefficient law \citep{claret00}, hereinafter referred to as claret-4,
\begin{equation}
\label{eqn:ld_law_claret4}
I_{\lambda}(\mu) = 1 - \sum_{k=1}^{4} a_n(1-\mu^{k/2}) ;
\end{equation}
\item a generalized $n^{\mbox{\footnotesize{th}}}$-degree polynomial law,
\begin{equation}
\label{eqn:ld_law_gen_poly}
I_{\lambda}(\mu) = 1 - \sum_{k=1}^{n} b_k(1-\mu^{k}) ;
\end{equation}
\item a generalized claret-$n$ law,
\begin{equation}
\label{eqn:ld_law_gen_claret}
I_{\lambda}(\mu) = 1 - \sum_{k=1}^{n} c_k(1-\mu^{k/2}) ;
\end{equation}
\end{enumerate}
Additionally, user-defined limb-darkening laws can be easily implemented. We recommend using the claret-4 law to achieve a model precision of $\lesssim$10~ppm in the analysis of exoplanetary transits (see Section~\ref{ssec:fitting_ld_laws}). The next release of \texttt{ExoTETHyS} will include a grid of white dwarf models, for which we have also found the claret-4 law to be significantly more accurate than the two-coefficient laws \citep{claret20}.

\subsubsection{From the passband-integrated intensities to the limb-darkening coefficients} \label{ssec:fit_ints}

The limb-darkening coefficients are obtained through a weighted least-squares fit of the passband-integrated intensity profile with weights proportional to the sampling interval in $r$, hereinafter referred to as \emph{weighted}-$r$ fit. The corresponding cost function is the weighted rms of residuals,
\begin{equation}
\label{eqn:w-rRMS}
\mbox{\emph{weighted}-}r \, \mbox{rms} = \left ( \frac{\sum_{i=1}^{n} w_i ( I_{\mbox{\footnotesize pass}} (\mu_i) - I_{\mbox{\footnotesize pass}}^{\mbox{\footnotesize law}} (\mu_i) )^2}{ \sum_{i=1}^{n} w_i } \right )^{\frac{1}{2}},
\end{equation}
with weights
\begin{equation}
\label{eqn:w-r_weights}
w_i = \begin{cases} (1-r_1) + 0.5 \, (r_1-r_2), & \mbox{if} \ i=1 \\
0.5 \, (r_{i-1}-r_{i+1}), & \mbox{if} \ 1<i<n\\
0.5 \, r_{n-1}, & \mbox{if} \ i=n \end{cases},
\end{equation}
where the $r_i$ are arranged in descending order, and $r_n =0$. The choice of cost function is optimized for the study of exoplanet transits, as detailed in Section~\ref{ssec:fitting_methods}. The performances of the spherical model fits are further enhanced by discarding those points with $r>0.99623$ (after rescaling as explained in Section~\ref{ssec:integ_ints}). This cut is a generalization of that implemented in the quasi-spherical (QS) fits by \cite{claret12}. For this reason, we rename the total fitting procedure explained here for the spherical intensity profiles as the \emph{weighted}-$r$ QS fit. Further details about the alternative fitting procedures are discussed in Section~\ref{ssec:fitting_methods}.

\subsubsection{Interpolation from the grid of stellar models}
\label{ssec:ldc_interp}

The process described in Sections~\ref{ssec:integ_ints}-\ref{ssec:fit_ints} enables the calculation of limb-darkening coefficients for the stellar-atmosphere models contained in the grid, starting from their precalculated specific intensities. The limb-darkening coefficients for an individual target with a generic set of stellar parameters are obtained by sequential linear interpolation through the following steps:
\begin{enumerate}
\item identification of the neighbors in the model-grid, i.e., the vertices of the cube in parameter space that contains the requested model (maximum 8 models);
\item calculation of the limb-darkening coefficients for each of the neighbors;
\item interpolation in $[M/H]$ between models with the same $T_{\mbox{\footnotesize{eff}}}$ and $\log{g}$, leading to a maximum of 4 sets of limb-darkening coefficients with the requested $[M/H]$;
\item interpolation in $\log{g}$ between the above calculated sets of coefficients with the same $T_{\mbox{\footnotesize{eff}}}$, leading to a maximum of 2 sets of limb-darkening coefficients with the requested $\log{g}$ and $[M/H]$;
\item interpolation in $T_{\mbox{\footnotesize{eff}}}$ between the above calculated sets of coefficients.
\end{enumerate}
We note that this sequential interpolation is possible because of the regularity of the model grids. 

\begin{figure*}[t]
\plotone{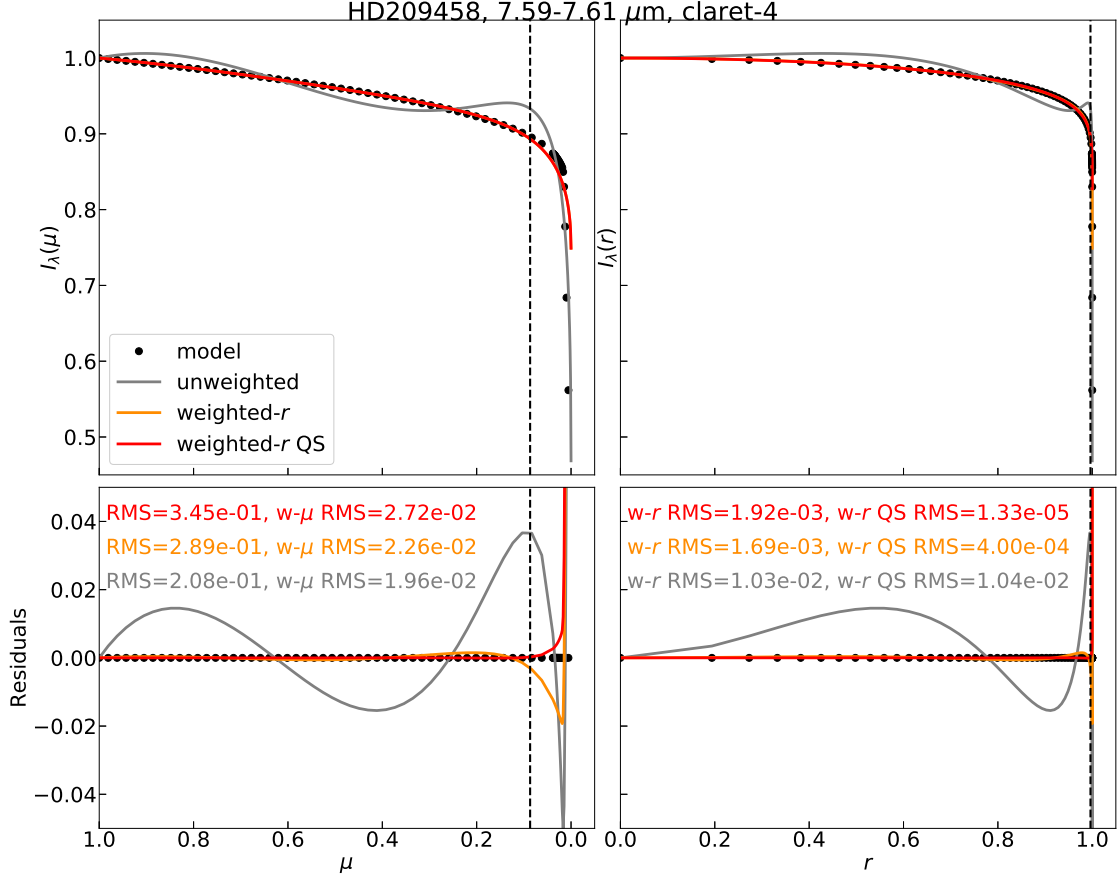}
\caption{Example with a model intensity distribution for a star similar to HD209458 ($T_{\mbox{\footnotesize eff}} = 6100 \, \mbox{K}$, $\log{g} = 4.5$), integrated over the 7.59--7.61~$\mu$m wavelength range, by using the \texttt{PHOENIX}\_2012\_13 database (see Table~\ref{tab:stellar_grids}). Top, left panel: normalized specific intensities vs. $\mu$ from the stellar atmosphere model (black circles), \emph{unweighted} (gray), \emph{weighted}-$r$ (orange), and \emph{weighted}-$r$ QS (red) model fits with claret-4 coefficients. The vertical dashed line denotes the cutoff value for the quasi-spherical fit (see Section~\ref{ssec:fitting_methods}). Top, right panel: analogous plot vs. $r$. Bottom panels: residuals between the fitted and model intensity values. The corresponding unweighted and weighted rms amplitudes of residuals are also reported. Note that, in this case, the unweighted least-squares fit leads to a non-monotonic radial intensity profile, which is physically unexpected. \label{fig:unphysical_ldfit}}
\end{figure*}

\begin{figure*}
\floatbox[{\capbeside\thisfloatsetup{capbesideposition={right,top},capbesidewidth=0.5\textwidth}}]{figure}[\FBwidth]
{\caption{Top panel: simulated transit light curve (black) of HD209458~b as it would be observed by \emph{TESS}, and best-fit model with claret-4 limb-darkening coefficients obtained with the \emph{weighted}-$r$ QS method (red). Bottom panels: residuals between the reference light curve and the best-fit models with claret-4 limb-darkening coefficients obtained with different limb-darkening laws and fitting methods (see Section~\ref{ssec:fitting_methods}). The peak-to-peak and rms amplitudes of the residuals are reported.}\label{fig:TESS_transits}}
{\hspace{-1.7cm}\includegraphics[width=0.37\textwidth]{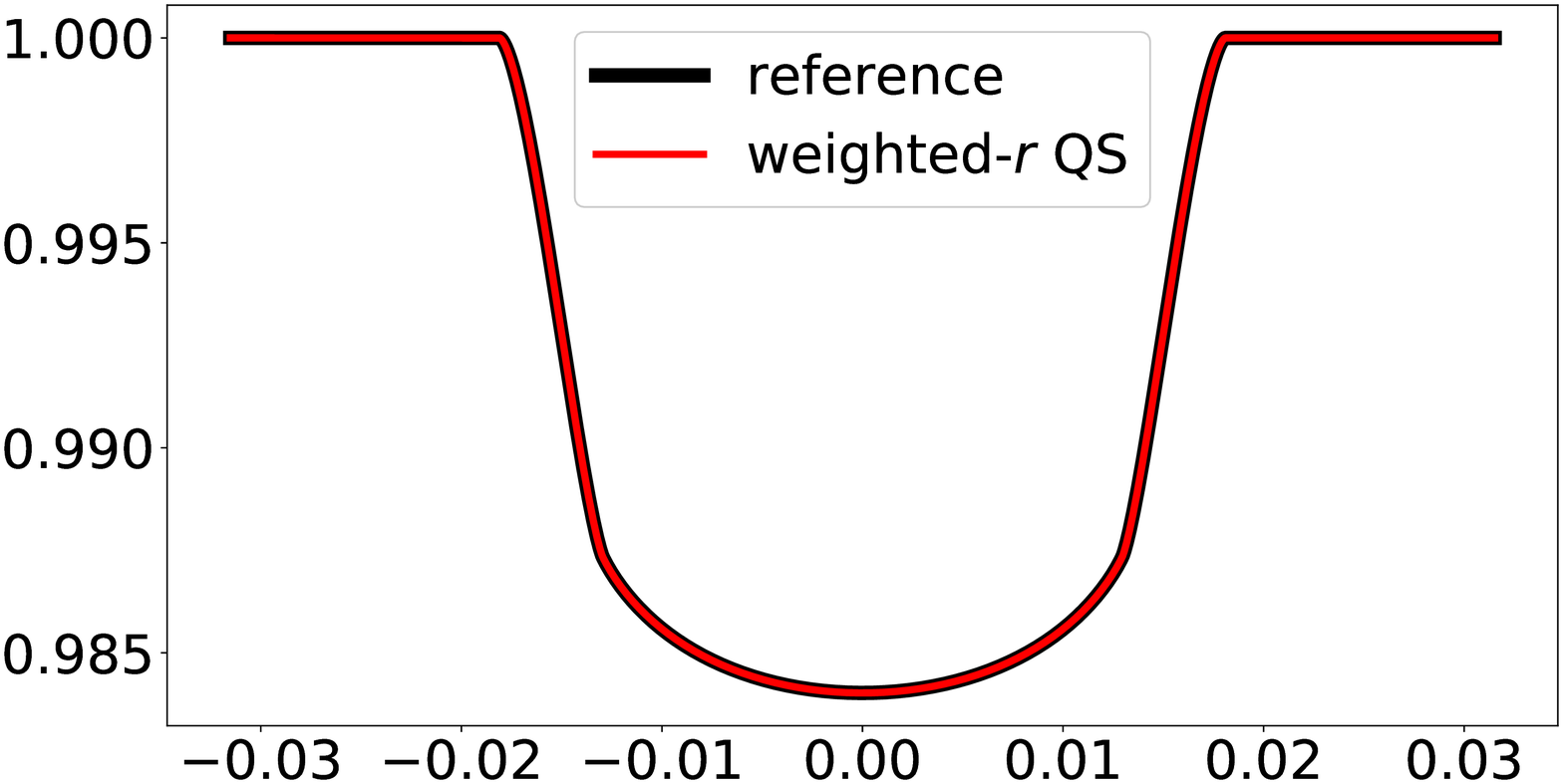}}
\includegraphics[width=\textwidth]{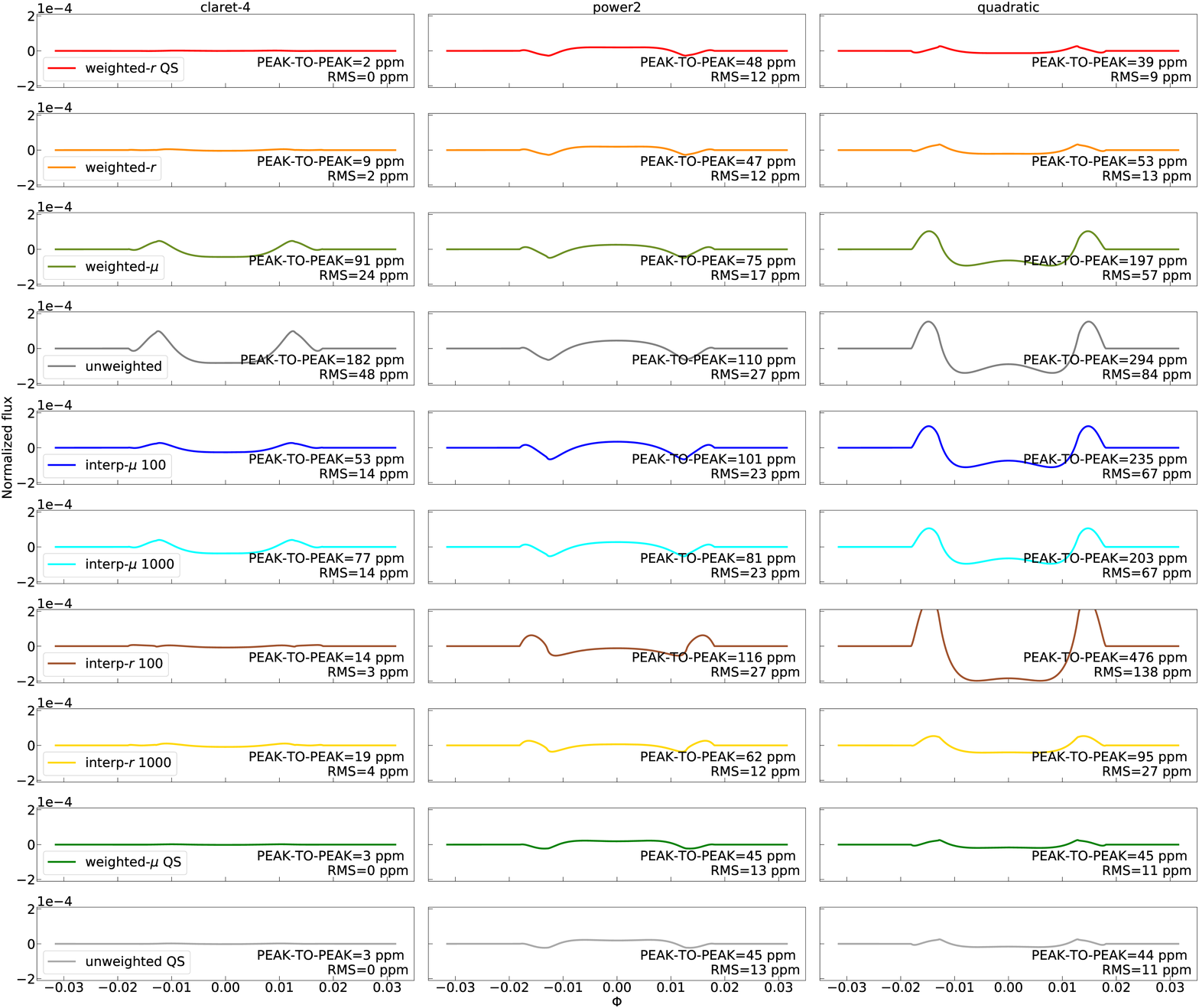}
\end{figure*}

\begin{figure*}[t]
\plotone{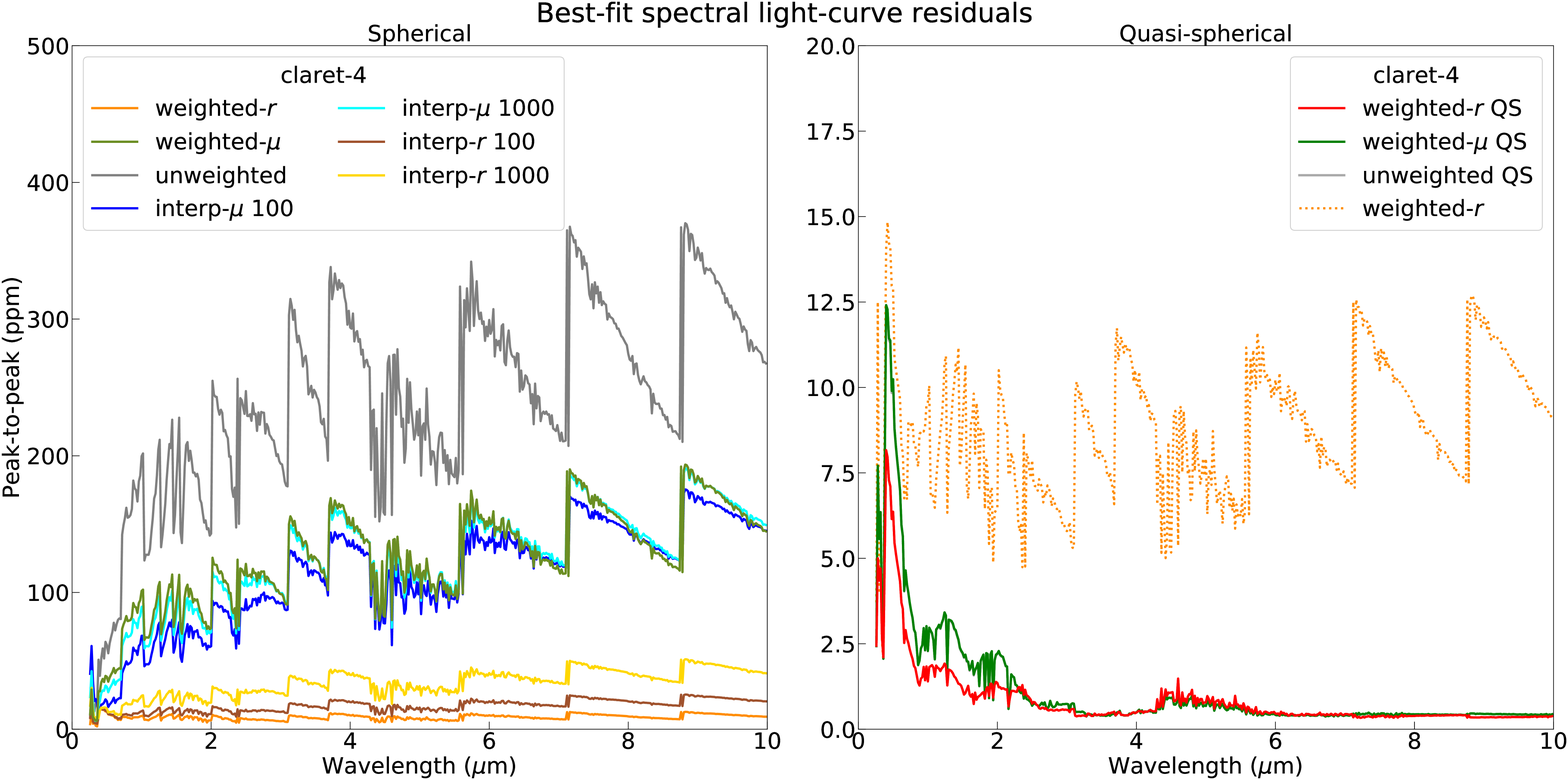}
\caption{Peak-to-peak of residuals between the reference spectral light curves for the transit of HD209458~b and the best-fit models with claret-4 limb-darkening coefficients obtained with different fitting methods (see Section~\ref{ssec:fitting_methods}). Left panel: results obtained with the spherical methods, i.e., taking into account the whole spherical intensity profiles. Right panel: results obtained with the quasi-spherical methods, i.e., with a cutoff of $r \le 0.99623$, and the \emph{weighted}-$r$ method (dotted, orange line). The \emph{unweighted} QS (gray line) and the \emph{weighted}-$\mu$ QS (green line) overlap in the plot. Note the scale difference between the two panels. \label{fig:spectrum_residuals}}
\end{figure*}

\subsection{The TRIP subpackage}
The TRIP subpackage is used to generate exact transit light curves by direct integration of the occulted stellar flux at given instants. It assumes a dark spherical planet transiting in front of a spherically-symmetric (unspotted) star. In this simple case, the normalized flux (i.e., relative to the stellar flux) is a function of two geometric variables, as reported by \cite{mandel02}, and of the stellar intensity profile:
\begin{equation}
\label{eqn:F_p_z}
F(p,z, I(\mu)) = 1 - \Lambda (p,z,I(\mu)) ,
\end{equation}
where $p$ is the planet-to-star radii ratio ($p=R_p/R_*$), $z$ is the sky-projected distance between the centers of the two spheres in units of the stellar radius, and $I(\mu)$ is the stellar intensity profile.
TRIP does not use an analytical approximation of the limb-darkening profile, unlike most transit light curve calculators such as those provided by \cite{mandel02}, \cite{gimenez06}, \cite{agol19}, \texttt{JKTEBOP} \citep{southworth04}, \texttt{TAP} \citep{gazak12}, \texttt{EXOFAST} \citep{eastman13},   \texttt{PyTransit} \citep{parviainen15}, \texttt{BATMAN} \citep{kreidberg15}, and \texttt{PYLIGHTCURVE}\footnote[8]{\url{https://github.com/ucl-exoplanets/pylightcurve}} \citep{tsiaras16}.  

\subsubsection{Input and output}
The TRIP subpackage requires a configuration file, where the user has to specify the name of the text files containing the limb-darkening profile, the phase, time, or $z$-series for which to calculate the normalized flux, and a list of parameter values that includes $p$ and those parameters eventually needed to compute the $z$-series (see Section~\ref{ssec:z_dist}). The limb-darkening file consists of two columns with the $\mu$ or $r$ values (first column) and the corresponding specific intensities (second column). A list of optional parameters can be used to set the calculation details, i.e., the number of annuli, the interpolation variable, and the polynomial order for the spline interpolation (see Section~\ref{ssec:norm_flux}). It is also possible to define simple operations on the original limb-darkening profile, i.e., a possible cutoff in $\mu$ or $r$ with or without rescaling the $\mu$ or $r$ values to the cutoff radius. The output is a text or pickle file containing the normalized flux series for the requested phase, time, or $z$-series.

\subsubsection{Computing the $z$-series} \label{ssec:z_dist}
In general, $z$ is a function of the orbital phase ($\Phi$), i.e., the fraction of orbital period ($P$) from the closest transit event:
\begin{equation}
\label{eqn:phi_definition}
\Phi = \frac{t - \mbox{E.T.}}{P} - n,
\end{equation}
where $t$ denotes time, E.T. is the Epoch of Transit (i.e., a reference mid-transit time), and $n$ is the number of orbits from the E.T. rounded to the nearest integer.
Conventionally, $\Phi$ values are in the range of $[-0.5, \ 0.5]$ or $[0, 1]$ and $\Phi=0$ at mid-transit time.
The projected star--planet separation is given by
\begin{equation}
\label{eqn:z_dist}
z = \begin{cases}
a_R \sqrt{1 - \cos^2{( 2 \pi \Phi )} \sin^2{i} } & \mbox{circular orbit} \\
a_R \frac{1-e^2}{1+e \cos{f}} \sqrt{1-\sin^2{(f+\omega)} \sin^2{i}} & \mbox{eccentric orbit}
\end{cases} ,
\end{equation}
where $a_R$ is the orbital semimajor axis in units of the stellar radius, $i$ is the inclination, $e$ is the eccentricity, $\omega$ is the argument of periastron, and $f$ is the true anomaly. In the eccentric case, the true anomaly is calculated from the orbital phase by solving Kepler's equation,
\begin{equation}
\label{eqn:kepler_ecc}
\frac{\pi}{2} - \omega + 2 \pi \Phi = E - e \sin{E} ,
\end{equation}
then
\begin{equation}
\label{eqn:true_anomaly}
f = 2 \arctan{ \left ( \sqrt{\frac{1+e}{1-e}} \tan{\frac{E}{2}} \right )} .
\end{equation}

\subsubsection{Calculating the normalized flux} \label{ssec:norm_flux}
The total and occulted stellar flux are given, respectively, by the integrals
\begin{equation}
\label{eqn:Fstar_integral}
F_{*} = \int_{0}^{1} I(r) \, 2 \pi r \, dr ,
\end{equation}
and
\begin{equation}
\label{eqn:Focc_integral}
F_{*,\mbox{\footnotesize occ}} = \int_{0}^{1} I(r) \, 2 \pi r \, f_{p,z}(r) \, dr ,
\end{equation}
with 
\begin{multline}
\label{eqn:fpzr_fraction}
f_{p,z}(r) =\\
\left.
\begin{cases}
\frac{1}{ \pi} \arccos{ \frac{r^2 + z^2 - p^2}{2zr}} & |z-p| < r < z+p \\
0 & r \le z-p \ \mbox{or} \  r \ge z+p \\
1 & r \le p-z
\end{cases} \right |_{0 \le r \le 1} .
\end{multline}
$I(r)$ is the specific intensity at the normalized radial coordinate $r=\sqrt{1-\mu^2}$, and $f_{p,z}(r)$ is the fraction of circumference with radius $r$ covered by the planet. Equations~\ref{eqn:Fstar_integral} and \ref{eqn:Focc_integral} rely on the assumed spherical symmetry for the star; Equation~\ref{eqn:fpzr_fraction} also makes use of the planet sphericity.
Finally, the normalized flux is given by Equation~\ref{eqn:F_p_z} with
\begin{equation}
\label{eqn:lambda_p_z}
\Lambda (p,z,I(\mu)) = \frac{F_{*,\mbox{\footnotesize occ}}}{F_{*}} .
\end{equation}
The integrals in Equations~\ref{eqn:Fstar_integral} and \ref{eqn:Focc_integral} are calculated numerically by using the mid-point rule with a uniform partition in $r$. The specific intensities are evaluated at the partition radii by interpolating in $\mu$ or $r$ from the input limb-darkening profiles. The TRIP algorithm with default settings is identical to the ``tlc'' described by \cite{morello17}.

\section{Performance of \texttt{ExoTETHyS}} \label{sec:performances}

\subsection{Comparison between fitting algorithms for the stellar intensity profiles} \label{ssec:fitting_methods}

\begin{figure*}[t]
\plotone{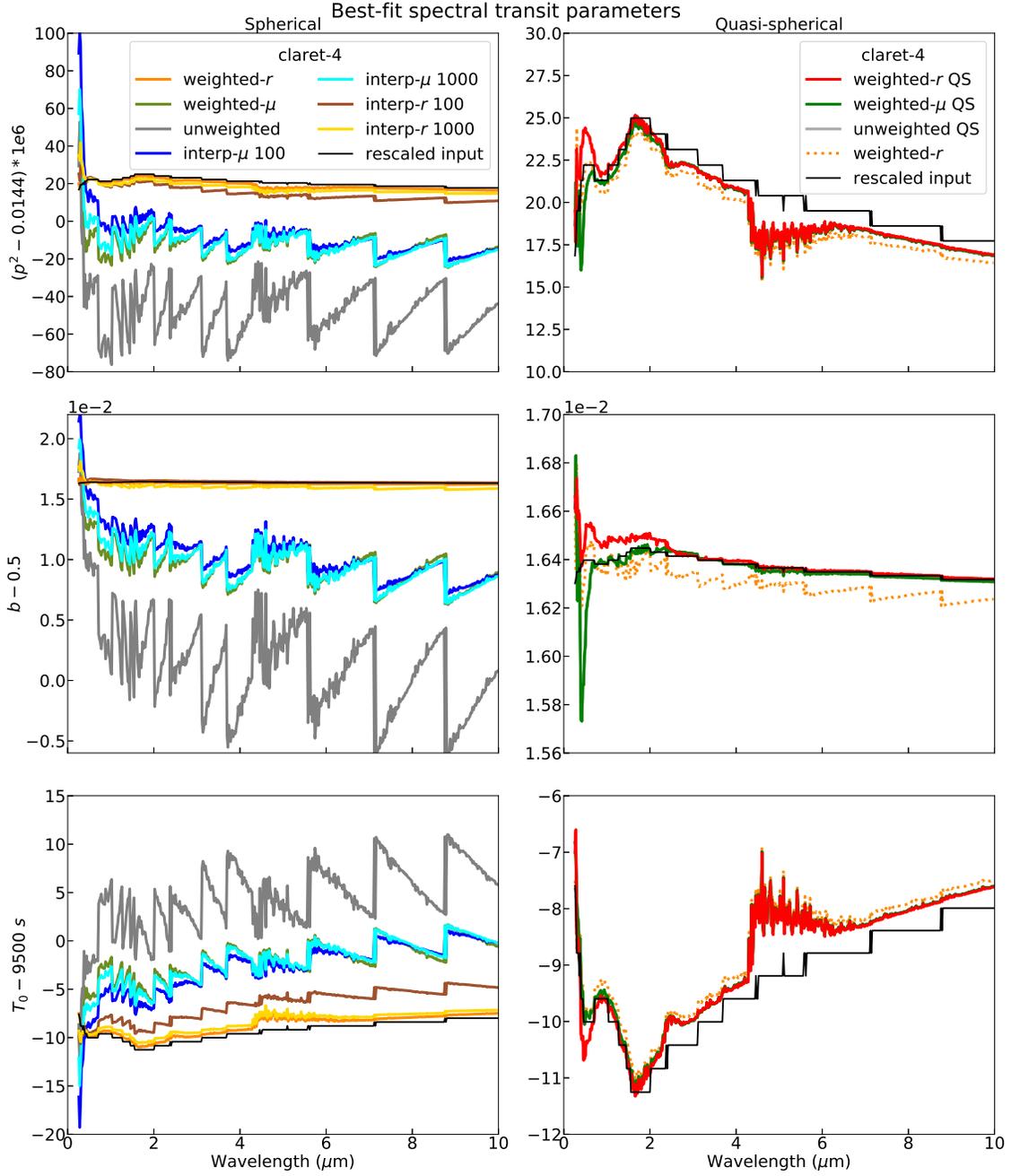}
\caption{Best-fit transit parameters to the reference spectral light curves for the transit of HD209458~b assuming claret-4 limb-darkening coefficients obtained with different fitting methods (see Section~\ref{ssec:fitting_methods}). The true parameter values are reported in black. Left panels: results obtained with the spherical methods, i.e., taking into account the whole spherical intensity profiles. Right panels: Results obtained with the quasi-spherical methods, i.e., with a cutoff of $r \le 0.99623$, and the \emph{weighted}-$r$ method (dotted, orange line). Note the scale difference between the two panels. \label{fig:spectrum_params}}
\end{figure*}

A long list of methods has been adopted in the literature for fitting the limb-darkening laws to the model intensity profiles leading to significantly different limb-darkening coefficients. The coefficients obtained with a simple least-squares fit depend on the spatial distribution of the precalculated intensities. The effect of sampling is particularly evident for the \texttt{PHOENIX}    profiles because of a much finer sampling near the drop-off region. For example, Figure~\ref{fig:unphysical_ldfit} shows the case of a star similar to HD209458 in the mid-infrared, for which the simple least-squares solution presents a non-monotonic (unphysical) profile with unexpected undulations.
In this paper, we compare the following fitting procedures:
\begin{enumerate}
\item \textit{unweighted}, i.e., simple least-squares fit;
\item \textit{weighted}-$r$, i.e., weighted least-squares fit with weights proportional to the sampling interval in $r$, as detailed in Equations~\ref{eqn:w-rRMS} and \ref{eqn:w-r_weights};
\item \textit{weighted-$\mu$}, i.e., weighted least-squares fit with weights proportional to the sampling interval in $\mu$;
\item \textit{interp-$\mu$~100}, i.e., least-squares fit on the intensities interpolated over 100 $\mu$ values with a uniform separation in $\mu$, as suggested by \cite{claret11};
\item \textit{interp-$\mu$~1000}, i.e., least-squares fit on the intensities interpolated over 1000 $\mu$ values with a uniform separation in $\mu$;
\item \textit{interp-$r$~100}, i.e., least-squares fit on the intensities interpolated over 100 $r$ values with a uniform separation in $r$, as suggested by \cite{parviainen15b} (with an unspecified number of interpolated values);
\item \textit{interp-$r$~1000}, i.e., least-squares fit on the intensities interpolated over 1000 $r$ values with a uniform separation in $r$;
\item \textit{unweighted} QS, i.e., least-squares fit with a cutoff $r \le 0.99623$;
\item \textit{weighted}-$r$ QS, i.e., analogous to \textit{weighted}-$r$ with a cutoff $r \le 0.99623$;
\item \textit{weighted}-$\mu$ QS, i.e., analogous to \textit{weighted}-$\mu$ with a cutoff $r \le 0.99623$.
\end{enumerate}
The cutoff is used to remove the steep drop-off characteristic of the spherical models, hence the term QS. The QS approach was first proposed by \cite{claret12}, who applied a cutoff $\mu \ge 0.1$ to their library of \texttt{PHOENIX} models with the original $\mu$ values. In this work, we redefine the cutoff using the rescaled $r$, such that it corresponds to the same fraction of the photometric stellar radius for all the models (see Section~\ref{ssec:integ_ints}). Our new definition with $r \le 0.99623$ is equivalent to the previous one for the majority of models, particularly for those models that may correspond to main-sequence stars. However, the libraries of \texttt{PHOENIX} models incorporated in the \texttt{ExoTETHyS} package also include models of stellar atmospheres with lower gravities than those analyzed by \cite{claret12}, corresponding to subgiant, giant, and supergiant stars. For some of these models, the intensity drop-off occurs at $\mu>0.1$, so that the cutoff of $\mu \ge 0.1$ (not rescaled) would be ineffective.

In order to evaluate the merits of the alternative fitting procedures to the stellar intensity profile, we generated exact synthetic transit light curves using the TRIP subpackage and compared these light curves with their best-fit solutions obtained with the various sets of claret-4 limb-darkening coefficients.
Figure~\ref{fig:TESS_transits} shows the residuals obtained for a noiseless simulation of the transit of HD209458~b in the \emph{TESS} passband when adopting the different sets of limb-darkening coefficients. The \emph{weighted}-$r$ QS method implemented in \texttt{ExoTETHyS}.SAIL gives the smallest residuals, with a peak-to-peak of 2~ppm and rms amplitude below 1~ppm. The other QS methods, \textit{weighted}-$\mu$ QS and \textit{unweighted} QS, lead to almost identical residuals, with a peak-to-peak of 3~ppm. Among the spherical methods, the \textit{weighted}-$r$ gives the smallest residuals with a peak-to-peak of 9~ppm and rms amplitude of 2~ppm, followed by the \textit{interp-$r$~100} and \textit{interp-$r$~1000} with about 1.5 and 2 times larger residual amplitudes, respectively. All the other methods lead to significantly larger residuals of tens to a few hundred ppm, which are comparable with the predicted noise floor of 60~ppm for the \emph{TESS} observations \citep{ricker14}.

Figure~\ref{fig:spectrum_residuals} shows the peak-to-peak of the residuals for the same transit as a function of wavelength, based on simulated light curves with 20~nm passband widths. This spectral analysis confirms the relative ranking of the fitting methods derived from the \emph{TESS} simulation. In particular, the \textit{weighted}-$r$ QS method leads to a peak-to-peak of residuals below 2~ppm at wavelengths longer than 1~$\mu$m, and overall below 8~ppm. The other quasi-spherical methods are marginally worse than \textit{weighted}-$r$ QS at wavelengths shorter than 2~$\mu$m, but the worst case peak-to-peak of residuals is less than 13~ppm. The \textit{weighted}-$r$ method leads to peak-to-peak of residuals in the range of 5-15~ppm, with a sawtooth-like modulation in wavelength. We noted that the small, but abrupt, jumps that occur at certain wavelengths correspond to changes of the inflection point in the stellar intensity profile as defined in Section~\ref{ssec:integ_ints}. The same phenomenon occurs for all the other spherical models with larger sawtooth-like modulations. It may appear surprising that the peak-to-peak of residuals obtained with the spherical methods tends to be larger at the longer wavelengths, for which the limb-darkening effect is expected to be smaller. The cause of the poor performances of most spherical methods in the infrared is the intensity drop-off, which is typically steeper than the drop-off in the UV and visible. Such drop-off has a negligible effect in the numerically integrated transit light curves, hence the better performances of the QS fits.

Figure~\ref{fig:spectrum_params} shows the best-fit transit parameters corresponding to the same spectral light curves, and compared with the respective input parameters corrected for the rescaled $r$ (see Section~\ref{ssec:integ_ints}). We retrieved the correct transit depth within 5~ppm, the impact parameter within 6$\times$10$^{-4}$, and the transit duration within 1~s at all wavelengths, when using the \textit{weighted}-$r$ or QS limb-darkening coefficients. However, slightly larger spectral trends appear in these parameters because of the wavelength-dependent stellar radius. The peak-to-peak variation in transit depth over the spectral range of 0.25--10~$\mu$m is 10~ppm. The other sets of limb-darkening coefficients introduce orders-of-magnitude larger biases in the retrieved transit parameters, also larger spectral sawtooth-like modulations in the infrared (few tens of ppm in transit depth across 1-10~$\mu$m), and severe discrepancies between the parameter values obtained in the UV/visible and those obtained in the infrared.

\begin{figure*}[t]
\plotone{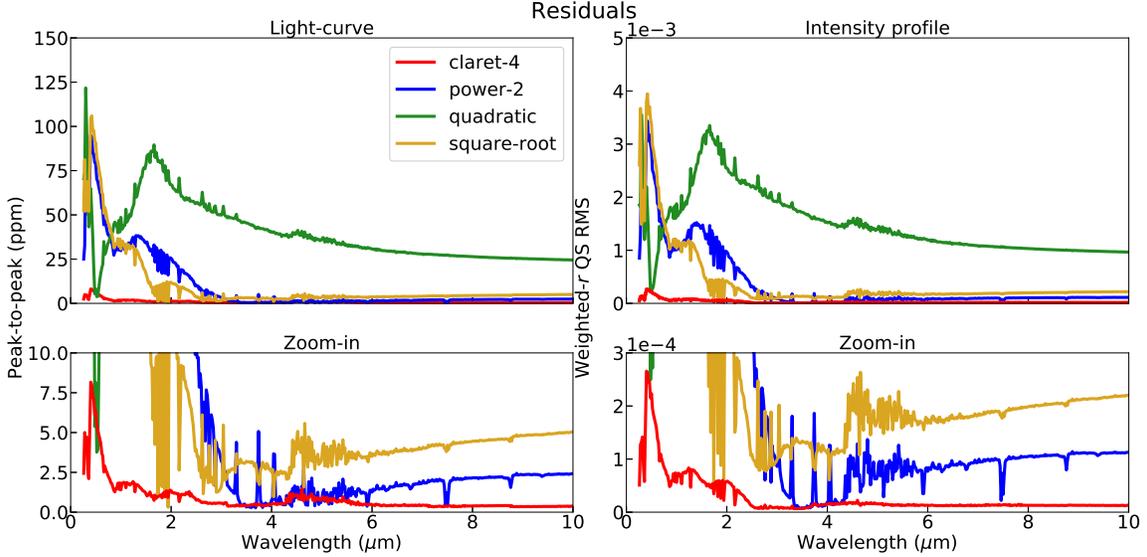}
\caption{Top, left panel: peak-to-peak of residuals between the reference spectral light curves for the transit of HD209458~b and the best-fit models using the limb-darkening coefficients calculated for the different laws (see Section~\ref{ssec:ld_laws}). Top, right panel: \emph{weighted}-$r$ QS rms of residuals to the model intensity profiles. Bottom panels: zoom-in of the panels above. \label{fig:spectrum_residuals_ldlaws}}
\end{figure*}

\begin{figure*}[t]
\plotone{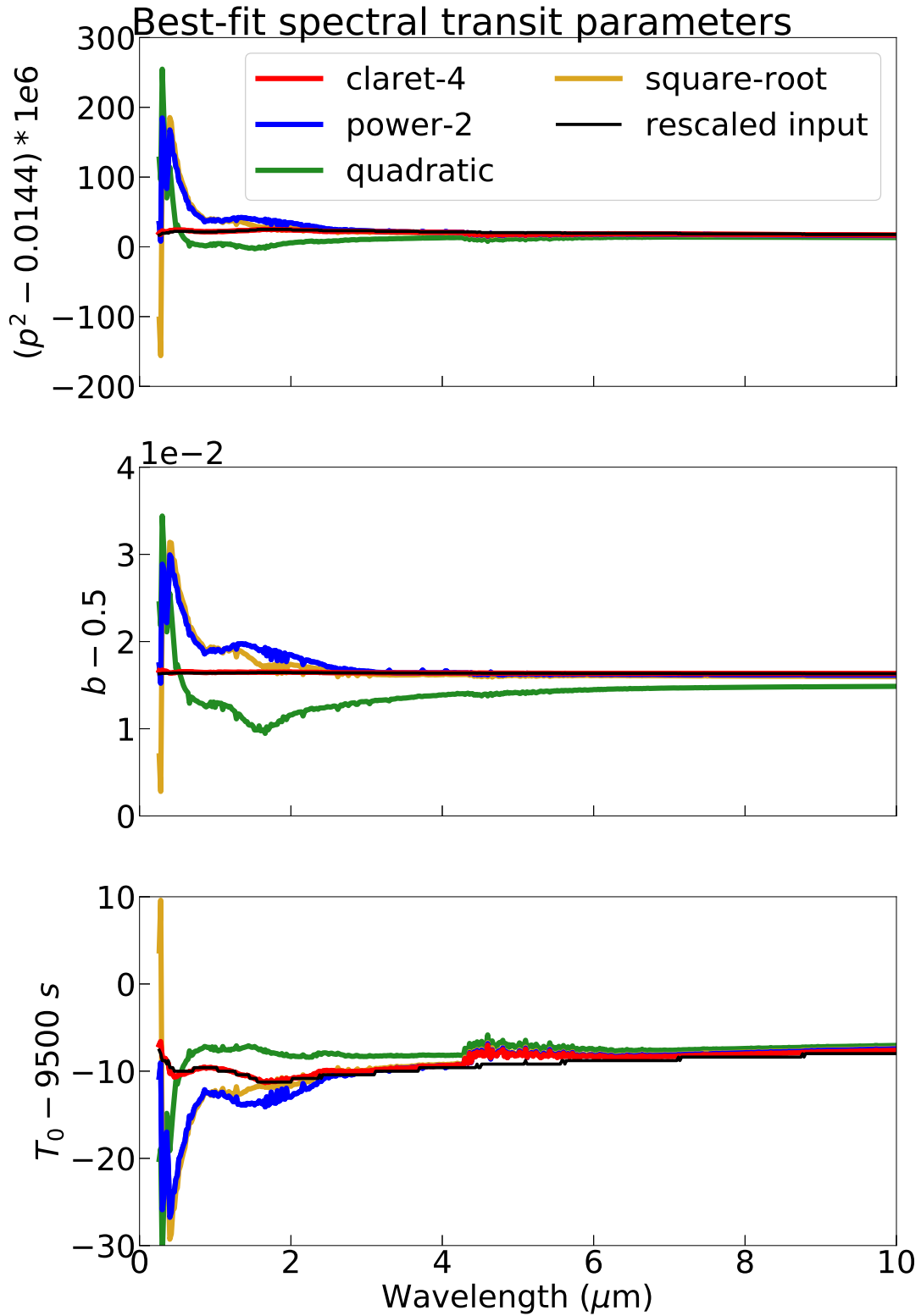}
\caption{Best-fit transit parameters to the reference spectral light curves for the transit of HD209458~b using the limb-darkening coefficients calculated for the different laws (see Section~\ref{ssec:ld_laws}). The true parameter values are reported in black. \label{fig:spectrum_params_ldlaws}}
\end{figure*}

\begin{table*}[t]
\centering
\caption{Spectral analysis of the error in transit depth when adopting different limb-darkening laws.} \label{tab:p2_bias}
\begin{tabular}{cccccc}
\tablewidth{0pt}
\hline
\hline
 & Wavelength range & Claret-4 & Power-2 & Quadratic & Square-root \\
\hline
Maximum bias & 0.25--10.0~$\mu$m & 5 & 165 & 235 & 174 \\
(ppm) & $<$1~$\mu$m & 4 & 165 & 235 & 174 \\
 & $>$1~$\mu$m & 5 & 19 & 27 & 18 \\
 & $>$5~$\mu$m & 3 & 4 & 10 & 5 \\
\hline
Rms bias & 0.25--10.0~$\mu$m & 1 & 20 & 20 & 23 \\
(ppm) & $<$1~$\mu$m & 2 & 71 & 62 & 81 \\
 & $>$1~$\mu$m & 1 & 5 & 11 & 4 \\
 & $>$5~$\mu$m & 1 & 2 & 6 & 2 \\
\hline
Spectrum & 0.25--10.0~$\mu$m & 10 & 177 & 258 & 341 \\
peak-to-peak & $<$1~$\mu$m & 7 & 177 & 254 & 341 \\
(ppm) & $>$1~$\mu$m & 10 & 27 & 17 & 25 \\
 & $>$5~$\mu$m & 2 & 3 & 4 & 3 \\
\hline
Spectrum std & 0.25--10.0~$\mu$m & 2 & 20 & 18 & 23 \\
(ppm) & $<$1~$\mu$m & 1 & 45 & 58 & 64 \\
 & $>$1~$\mu$m & 2 & 7 & 4 & 5 \\
 & $>$5~$\mu$m & $<$1 & $<$1 & $<$1 & $<$1 \\
\hline
\end{tabular}
\end{table*}

\subsection{Performance of the limb-darkening laws}
\label{ssec:fitting_ld_laws}
Figure~\ref{fig:spectrum_residuals_ldlaws} compares the peak-to-peak of the spectral light curve residuals when adopting the limb-darkening coefficients calculated by \texttt{ExoTETHyS}.SAIL for different limb-darkening laws, as well as the corresponding \emph{weighted}-$r$ QS rms of the residuals to the stellar intensity profiles. The correlation between the two goodness-of-fit measures is explored in Section~\ref{ssec:GOF_ppm}.
At wavelengths $\gtrsim$3~$\mu$m, the precision of the power-2 and square-root limb-darkening coefficients is comparable to that of the claret-4 coefficients, resulting in light curve residuals below 5~ppm. While the claret-4 law performs similarly well even at shorter wavelengths, the two-coefficient laws lead to larger light curve residuals up to $\sim$100~ppm in the UV and visible. The quadratic law is less precise, leading to light curve residuals above 25~ppm even at 10~$\mu$m.

Figure~\ref{fig:spectrum_params_ldlaws} shows the fitted transit parameters and their expected values. Typically, the bias in transit depth is of the same order of magnitude of the light curve residuals, but it can be both larger or smaller than their peak-to-peak amplitudes owing to parameter degeneracies. Table~\ref{tab:p2_bias} reports the statistics of the errors in transit depth obtained with the different limb-darkening laws across given spectral ranges. The maximum bias in transit depth at 5--10~$\mu$m is within 10~ppm for any limb-darkening parameterization, which is just below the minimum photon noise floor for \emph{JWST}/Mid-InfraRed Instrument (MIRI) observations \citep{beichman14}. At $\sim$1~$\mu$m, the two-coefficient laws may introduce a spectral slope of a few tens of ppm, which may have an impact in the analysis of the \emph{HST}/WFC3 spectra \citep{tsiaras18}. At wavelengths shorter than 1~$\mu$m the two-coefficient laws are unreliable for exoplanet spectroscopy, so that the claret-4 law must be preferred.
These conclusions are in agreement with previous studies based on both simulated and real data \citep{espinoza16, morello17, morello18, maxted18}.

\begin{figure}[t]
\plotone{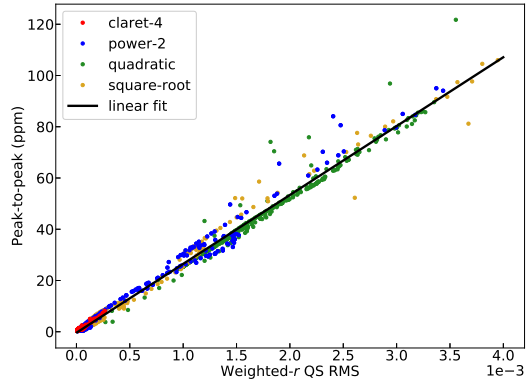}
\caption{\emph{Weighted}-$r$ QS rms of residuals to the model intensity profiles vs. peak-to-peak of the transit light curve residuals for the spectral templates of HD20458~b adopting different limb-darkening laws. The black line is the global linear fit.  \label{fig:resints_vs_reslc}}
\end{figure}

\subsection{Predicted precision in light curves}
\label{ssec:GOF_ppm}
Figure~\ref{fig:resints_vs_reslc} shows that, for a fixed transit geometry, the peak-to-peak of light curve residuals is roughly proportional to the \emph{weighted}-$r$ QS rms of stellar intensity residuals. We found an approximately linear correlation between the two goodness-of-fit measures for the simulated spectral light curves and stellar intensity profiles, therefore obtaining a wavelength-independent factor. We repeated this test for analogous sets of spectral light curves with different transit parameters, then obtaining different proportionality factors. Our preliminary study suggests that
\begin{multline}
\label{eqn:GOF_prop}
(\mbox{peak-to-peak})_{\mbox{\footnotesize ppm}} = (k \times 10^6) \times p^2 \\
\times (\mbox{\emph{weighted}-}r \, \mbox{QS} \, \mbox{rms} ) ,
\end{multline}
where $k$ is a factor of order unity ($k \gtrsim 1$).
Equation~\ref{eqn:GOF_prop} provides a useful tool for estimating the systematic noise in the light curve models solely due to the limb-darkening parameterization. The systematic noise in the light curve models should be smaller than the photon noise limit of the observation in order to avoid significant parameter biases. Note that Equation~\ref{eqn:GOF_prop} does not account for uncertainties in the stellar parameters, discrepancies between real and model intensity profiles, and other contaminating signals that may increase the total systematic noise.

\section{Usage of \texttt{ExoTETHyS}} \label{sec:usage}

Currently, the main use of the \texttt{ExoTETHyS} package is to compute stellar limb-darkening coefficients through the SAIL subpackage. These coefficients can be adopted to simulate the exoplanetary transit light curves, which are largely used by the scientific consortia of the future exoplanet missions for multiple studies. In particular, \texttt{ExoTETHyS} will be linked with \emph{ARIEL}-Sim \citep{sarkar16} and \texttt{ExoNoodle} (a generator of timeseries spectra of exoplanetary systems originally designed for \emph{JWST} observations; M. Martin-Lagarde et al., in prep.), and it has already been adopted by several members of the two mission consortia.

It is also common practice to fix the limb-darkening coefficients obtained from stellar models, such as those calculated with \texttt{ExoTETHyS}.SAIL, in the analysis of exoplanetary transit light curves. This approach relies on the perfect match between the model and the real stellar intensity distributions, otherwise introducing a potential bias in the derived exoplanet and orbital parameters. Some authors recommended setting free limb-darkening coefficients in the light curve fits to minimize the potential bias, but the strong parameter degeneracies may lead to larger error bars or prevent the convergence of the fit \citep{southworth08, csizmadia13}.
The parameter degeneracies can be mitigated by using multiwavelength transit observations to better constrain the orbital parameters \citep{morello17, morello18}.
Here we suggest an approach to take advantage of the knowledge on the stellar parameters in the form of bayesian priors. The stellar parameters will then be optimized in the light curve fits instead of using fixed or fully unconstrained limb-darkening coefficients. The limb-darkening coefficients for a given set of stellar parameters, and a given passband or spectroscopic bin, can be interpolated from a precalculated grid. The grid calculation type (see Section~\ref{ssec:sail_IO}) was specifically designed for this purpose.

\section{Conclusions} \label{sec:conclusions}
We introduced \texttt{ExoTETHyS}, an open-source python package that offers accessory tools for modeling transiting exoplanets and eclipsing binaries. It includes a versatile stellar limb-darkening calculator with multiple choices of model atmosphere grids, parameterizations, passbands (also accepting user input), and specific user-friendly calculation settings. We demonstrated an optimal fitting algorithm for the limb-darkening coefficients, thus eliminating the degree of freedom associated with the choice of fitting algorithm. The claret-4 coefficients obtained through this algorithm ensure a precision level $\lesssim$10~ppm in the relevant transit light curves at all wavelengths. The precision achieved exceeds by one order of magnitude that obtained with most of the algorithms proposed in the previous literature for stellar models with spherical geometry. We also proposed a simple formula for estimating the light curve model precision, based on the goodness-of-fit for the limb-darkening coefficients.
Finally, we discussed the current and future usage of \texttt{ExoTETHyS} with emphasis on exoplanet atmospheric spectroscopy in the era of  \emph{JWST} and ARIEL.



\acknowledgments
The authors would like to thank Ren{\'e} Gastaud and Daniel Dicken for useful discussions.
G. M., M. M.-L. and P.-O. L. were supported by the LabEx P2IO, the French ANR contract 05-BLANNT09-573739 and the European Unions Horizon 2020 Research and Innovation Programme, under Grant Agreement N$^\circ$776403. G.M. also acknowledges the contribution from INAF through the ``Progetto Premiale: A Way to Other Worlds'' of the Italian Ministry of Education, University, and Research. 
A.C. acknowledges financial support from the Spanish MEC (AYA2015-71718-R and ESP2017-87676-C5-2-R), the State Agency for Research of the Spanish MCIU through the ``Center of Excellence Severo Ochoa'' award for the Instituto de Astrof\'isica de Andaluc\'ia (SEV-2017-0709).


\bibliographystyle{aasjournal}
\bibliography{mybib}



\end{document}